# Extended plane wave expansion formulation for viscoelastic phononic thin plates

E.J.P. Miranda Jr.[a,b,c,*], V.F. Dal Poggetto[d], N.M. Pugno[d,e], J.M.C. Dos Santos[f]

[a]*Federal Institute of Maranhao, IFMA-EIB-DE, Rua Afonso Pena, 174, CEP 65010-030, São Luís, MA, Brazil.*
[b]*Federal Institute of Maranhão, IFMA-PPGEM, Avenida Getúlio Vargas, 4, CEP 65030-005, São Luís, MA, Brazil.*
[c]*Vale Institute of Technology, ITV-MI, Rua Professor Paulo Magalhães Gomes - Bauxita, CEP 35400-000, Ouro Preto, MG, Brazil.*
[d]*Laboratory for Bio-inspired, Bionic, Nano, Meta Materials & Mechanics, Department of Civil, Environmental and Mechanical Engineering, University of Trento, 38123 Trento, Italy.*
[e]*School of Engineering and Materials Science, Queen Mary University of London, Mile End Road, London E1 4NS, United Kingdom.*
[f]*University of Campinas, UNICAMP-FEM-DMC, Rua Mendeleyev, 200, CEP 13083-970, Campinas, SP, Brazil.*

**Abstract**

The extended plane wave expansion (EPWE) formulation is derived to obtain the complex band structure of flexural waves in viscoelastic thin phononic crystal plates considering the Kirchhoff-Love plate theory. The presented formulation yields the evanescent behavior of flexural waves in periodic thin plates considering viscoelastic effects. The viscosity is modeled by the standard linear solid model (SLSM), typically used to closely model the behavior of polymers. It is observed that the viscoelasticity influences significantly both the propagating and evanescent Bloch modes. The highest unit cell wave attenuation of the viscoelastic phononic thin plate is found around a filling fraction of 0.37 for higher frequencies considering the least attenuated wave mode. This EPWE formulation broadens the suitable methods to handle evanescent flexural waves in 2-D thin periodic plate systems considering the effects of viscoelasticity on wave attenuation.

[*]Corresponding author. Tel.: +55 9832226350
*Email address:* edson.jansen@ifma.edu.br (E.J.P. Miranda Jr.)

---

**1. Introduction**

Phononic structures (PnSs) are artificial composite materials composed of unit cells arranged in a specific spatially periodic form [1–3]. By tailoring the material composition and/or the spatial arrangement of the unit cell, the PnSs exhibit unusual band structure characteristics, such as Bragg scattering band gap formation. They have been also applied to vibration reduction [4–6], wave manipulation [7], energy harvesting [8, 9], as mechanical wave filters [10], seismic wave shields [11], and acoustic barriers [12, 13], among others.

The wave propagation characteristics of PnSs have been widely investigated for 1-D [14], 2-D [15–21], and 3-D [22, 23] cases. A fairly common type of investigated 2-D PnS is the phononic crystal considering thin plate theories [15–21], since the plate structures are widely used in aeronautical, mechanical and civil engineering, aerospace, and manufacturing applications [24]. Even though the propagating behavior of flexural waves in 2-D phononic thin plates has already been reported [15–21] by using a $\omega(\mathbf{k})$ (where $\omega$ is the angular frequency and $\mathbf{k}$ is the Bloch wave vector) approach (*i.e.*, considering propagating frequencies and neglecting the evanescent wave behavior), it should be highlighted that the flexural evanescent wave behavior (*i.e.,* both the real and imaginary parts of the wave vector, respectively, $\Re\{\mathbf{k}(\omega)\}$ and $\Im\{\mathbf{k}(\omega)\}$) has not been reported yet for 2-D phononic thin plates. As a result, the unit cell wave attenuation (*i.e.*, $\Im\{\mathbf{k}(\omega)\}a$, where $a$ is the lattice parameter), remains not investigated for flexural waves in 2-D phononic thin plates. This, in turn, is associated with the difficulty to formulate a $\mathbf{k}(\omega)$ approach (*i.e.*, considering any value of frequency and computing complex values for the wavevector $\mathbf{k}$) to obtain the flexural evanescent modes.

It is important to mention that in 2000s, some authors [25–28] reported very interesting results about flexural waves (*i.e.*, propagating [25, 27] and evanes-cent [26, 28] waves) in 2-D phononic thin plates. However, only $||\mathbf{k}(\omega)||$ and $\Re\{\omega(\mathbf{k})\}$ were computed and the complex band structure, typically obtained in terms of both $\Re\{\mathbf{k}(\omega)\}$ and $\Im\{\mathbf{k}(\omega)\}$ in the first Brillouin [29] zone (FBZ) for 2-D PnSs [30–33], was not reported. Poulton *et al.* [27] presented converged band structures ($\Re\{\omega(\mathbf{k})\}$) for Bloch-Floquet bending waves in a phononic thin plate containing a square array of circular inclusions, using the multipole formulation and applied in the situation where the perforations are no longer considered to be small in comparison with the lattice pitch. Movchan and collaborators [28] presented an analytical approach to model the Bloch-Floquet waves ($||\mathbf{k}||$) in structured Mindlin plates. They performed a comparative analysis of two simplified plate models, that is the classical Kirchhoff-Love theory and the Mindlin theory for dynamic response of periodic structure.

In the context of wave propagation in PnSs, an important issue that is sometimes commonly neglected in the complex band structure calculations is the viscoelastic effect present in many components, such as polymers. Moreover, the analysis of viscoelasticity and band structure of periodic structures has become an interesting topic for both the mathematical community [34–37] and engineering applications [33, 38]. The viscoelastic effect on the evanescent Bloch waves was firstly reported by [39, 40] for PnSs in a plane strain condition (*i.e.*, with infinite thickness). To the best of our knowledge, the influence of viscoelasticity on the complex band structure of 2-D phononic thin plates, considering the Kirchhoff-Love [41, 42] plate theory and only flexural waves, has not been studied.

The extended plane wave expansion (EPWE) method is a semi-analytical $\mathbf{k}(\omega)$ approach which has been extensively used to compute the complex band structure of 2-D acoustic metamaterials [14, 24, 43–45] and PnSs [30–32, 46] since this approach presents similar result as methods based on finite elements, but with a considerably lower computational cost [20]. The EPWE can obtain both propagating (purely real values of $\mathbf{k}$) and evanescent (imaginary and/or complex conjugate values of $\mathbf{k}$) wave modes. It should be highlighted that the wave modes computed by the EPWE are not restricted to the FBZ [30].

However, Hsue *et al.* [47] proved that the evanescent modes obtained by the EPWE obey Floquet-Bloch's theorem [48, 49]. One limitation of plane wave expansion (PWE) and EPWE methods is that both approaches can handle only infinite structures, *i.e.*, only the band structure can be obtained.

There are few previous studies that focused on phononic structures with viscoelastic components whose band structure was computed using the PWE [50, 51] and EPWE [37, 38, 40] approaches. Zhao and Wei [50, 51] computed the band structure (using the PWE) of 1-D and 2-D solid (infinite thickness) phononic crystals with viscoelasticity modelled by the standard linear solid model (SLSM). However, they did not obtain the evanescent waves, *i.e.*, omitting the information of the unit cell wave attenuation. Moiseyenko and Laude [40] calculated the evanescent wave modes for 2-D solid (infinite thickness) phononic crystals with the simple Kelvin-Voigt model. Thus, the complex band structure cannot handle the viscosity in a more realistic way, since the Kelvin-Voigt model is limited.

The main purpose of this study is to derive the EPWE formulation to compute the complex band structure of flexural waves propagating in a phononic thin plate using the Kirchhoff-Love theory [41, 42], with square inclusions distributed in a square lattice and the presence of viscoelastic effects. Viscosity is modelled by the SLSM, which contains three elements, *i.e.*, a Maxwell model (a spring and dashpot in series) and a spring in parallel [52].

The paper is organized as follows. Section 2 presents the new EPWE formulation for a phononic thin plate considering viscoelastic effects based on Kirchhoff-Love plate theory [41, 42]. In Section 3, a numerical example is carried out. Conclusions are presented in Section 4.

## 2. Viscoelastic Phononic Thin Plate Modelling

This section describes the EPWE formulation for a phononic Kirchhoff-Love [41, 42] thin plate considering viscoelastic effects. We consider wave propagation in the *xy* plane in a 2-D periodic isotropic medium.

The EPWE formulation is derived to investigate the evanescent flexural waves in phononic thin plates considering the SLSM. The SLSM contains three elements, *i.e.*, a Maxwell model (a spring and dashpot in series) and a spring in parallel. The viscosity presents a more realistic behavior involving a single exponential term in both creep and relaxation [52]. Figure 1 (*a*) sketches the top view of the 2-D phononic thin plate containing square inclusions in a square lattice. The phononic thin plate is composed by hard elastic inclusions and a soft viscoelastic matrix.

The SLSM [52] (Fig. 1 (*b*)) is used to consider the viscoelasticity of the material that forms the soft matrix, where $G_1$ and $G_2$ are the shear modulus (springs) and $\eta$ is the viscosity (dashpot) [52]. It should be highlighted that the SLSM has two forms, *i.e.*, the Maxwell and Kelvin forms. In this study, the Maxwell form is used, and the term SLSM means the Maxwell form of the SLSM [53]. In Fig. 1 (*c*), it is illustrated the first irreducible Brillouin [29] zone (FIBZ) for a square lattice (*a*), where $k_1, k_2 \in \mathbb{R}$ are the point coordinates within the FIBZ, $\bar{\varphi}$ is the azimuth angle of **k**, and the FIBZ high-symmetry points are $\Gamma$ $(0, 0)$, X $(\pi/a, 0)$, and M $(\pi/a, \pi/a)$.

## 2.1. Extended Plane Wave Expansion

The governing equation for the flexural vibration of a uniform isotropic thin plate considering the Kirchhoff-Love model [41, 42] composed by material *B* (see Fig. 1 (*a*)) without viscoelastic components can be written in the spatiotemporal domain as:

$$-a_B \ddot{\hat{w}}(\mathbf{r}, t) = \frac{\partial^2}{\partial x^2} \left[ D_B \frac{\partial^2 \hat{w}(\mathbf{r}, t)}{\partial x^2} + \beta_B \frac{\partial^2 \hat{w}(\mathbf{r}, t)}{\partial y^2} \right] + 2 \frac{\partial^2}{\partial x \partial y} \left[ \gamma_B \frac{\partial^2 \hat{w}(\mathbf{r}, t)}{\partial x \partial y} \right] + \frac{\partial^2}{\partial y^2} \left[ D_B \frac{\partial^2 \hat{w}(\mathbf{r}, t)}{\partial y^2} + \beta_B \frac{\partial^2 \hat{w}(\mathbf{r}, t)}{\partial x^2} \right], \quad (1)$$

where $a_B = \rho_B h$, $\rho_B$ is the material specific mass density, $h$ is the plate thickness, $D_B = E_B h^3/12(1 - \nu_B^2)$ is the plate flexural stiffness, $E_B$ is the material Young's modulus, $\nu_B$ is the material Poisson's ratio, $\beta_B = D_B \nu_B$, $\gamma_B = D_B(1 - \nu_B)$, $\hat{w}(\mathbf{r}, t)$ is the transverse displacement, *t* is the time, $\mathbf{r} = x\mathbf{e}_1 + y\mathbf{e}_2$ ($x, y \in R$) is the two-dimensional spatial vector, and $\mathbf{e}_i$ ($i$ = 1, 2) are the basis vectors of the periodic lattice in the real space. To facilitate the mathematical notation, hereafter the indexes $A$ and $B$ are related to elastic hard inclusions and viscoelastic soft matrix of the viscoelastic phononic thin plate (Fig. 1 (a)), respectively.

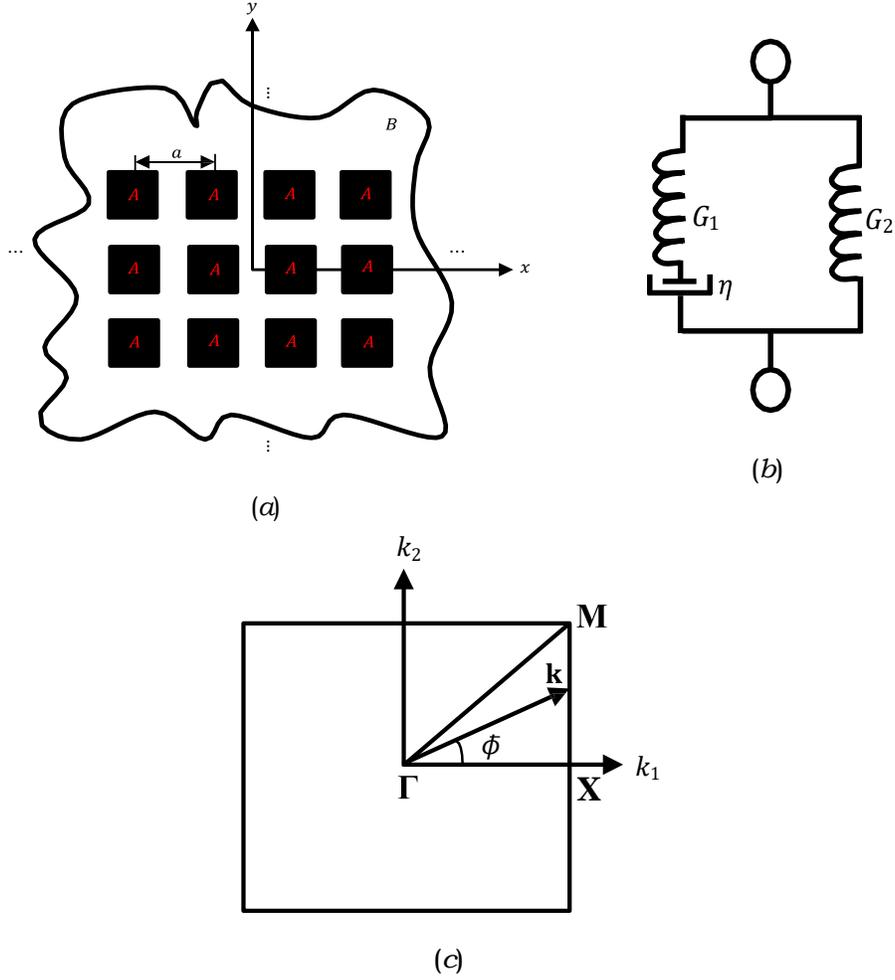

Figure 1: (a) Top view of the infinite 2-D viscoelastic PnS plate with square hard inclusions (A) in a soft matrix (B) with a square lattice, where $a$ is the lattice parameter, (b) the SLSM (where $\eta$ is the viscosity and $G_{1,2}$ is the shear modulus), and (c) the first irreducible Brillouin zone (FIBZ) (where $k_1$, $k_2 \in R$ are the point coordinates within the FIBZ, $\bar{\varphi}$ is the azimuth angle of **k**, $\Gamma$ (0, 0), X ($\pi/a$, 0), and M ($\pi/a$, $\pi/a$) are the high-symmetry points).

Equation (1) can also be rewritten for a uniform viscoelastic isotropic Kirchhoff-Love thin plate [41, 42] (composed by material $B$, see Fig. 1 ($a$)), however, it should be revisited that the constitutive equations, in the spatiotemporal domain, for a linearly viscoelastic material are given by [54, 55]:

$$\hat{\sigma}_{ij}(t) = \int_{-\infty}^{t} \hat{c}_{ijkl_B}(t-\tau) \frac{d\hat{E}_{kl}(t)}{d\tau} d\tau, \quad (2)$$

where $\{i, j, k, l\} = 1, 2, 3$ refer to the tensor indices, $\hat{\sigma}_{ij}$ is the elastic stress tensor, $\hat{c}_{ijkl_B}$ is the elastic stiffness tensor, $\hat{E}_{ij}$ is the elastic strain tensor, and $\tau$ is a time constant. The standard tensor notation is used with Latin indices running from 1 to 3, obeying Einstein's summation convention when repeated. The integration in Eq. (2) is known as the Boltzmann [52] or a hereditary integral, which expresses a convolution.

Applying the temporal Fourier transform to Eq. (1) and considering Eq. (2), thus, the governing equation for flexural vibration of a uniform viscoelastic isotropic Kirchhoff-Love thin plate [41, 42] can be rewritten as:

$$\frac{\partial^2}{\partial x^2}\left[i\omega D_B(\omega)\frac{\partial^2 w(\mathbf{r},\omega)}{\partial x^2} + i\omega \beta_B(\omega)\frac{\partial^2 w(\mathbf{r},\omega)}{\partial y^2}\right] + 2\frac{\partial^2}{\partial x \partial y}\left[i\omega \gamma_B(\omega)\frac{\partial^2 w(\mathbf{r},\omega)}{\partial x \partial y}\right]$$
$$+ \frac{\partial^2}{\partial y^2}\left[i\omega D_B(\omega)\frac{\partial^2 w(\mathbf{r},\omega)}{\partial y^2} + i\omega \beta_B(\omega)\frac{\partial^2 w(\mathbf{r},\omega)}{\partial x^2}\right] - \omega^2 a_B w(\mathbf{r},\omega) = 0, \quad (3)$$

where $D_B(\omega)$, $\beta_B(\omega)$, $\gamma_B(\omega)$, and $w(\mathbf{r},\omega)$ are the Fourier transforms of $\hat{D}_B(t)$, $\hat{\beta}_B(t)$, $\hat{\gamma}_B(t)$, and $\hat{w}(\mathbf{r}, t)$, respectively, and $\omega$ is the angular frequency. We notice that the term $i\omega$ multiplying the elastic constants ($D_B(\omega)$, $\beta_B(\omega)$, $\gamma_B(\omega)$) in Eq. (3) comes from the temporal derivative in Eq. (2). It is important to highlight that $\nu$ is considered approximately as a constant in this case, an assumption also considered in previous studies [56].

Regarding the viscoelastic phononic thin plate described in Fig. 1 ($a$), there are two materials, *i.e.*, a soft (viscoelastic) matrix and hard (elastic) inclusions, thus one can rewrite Eq. (3) as:

$$\frac{\partial^2}{\partial x^2}\left[D(\mathbf{r}, \omega)\frac{\partial^2 w(\mathbf{r}, \omega)}{\partial x^2} + \beta(\mathbf{r}, \omega)\frac{\partial^2 w(\mathbf{r}, \omega)}{\partial y^2}\right] + 2\frac{\partial^2}{\partial x \partial y}\left[\gamma(\mathbf{r}, \omega)\frac{\partial^2 w(\mathbf{r}, \omega)}{\partial x \partial y}\right]$$
$$+ \frac{\partial^2}{\partial y^2}\left[D(\mathbf{r}, \omega)\frac{\partial^2 w(\mathbf{r}, \omega)}{\partial y^2} + \beta(\mathbf{r}, \omega)\frac{\partial^2 w(\mathbf{r}, \omega)}{\partial x^2}\right] - \omega^2 a(\mathbf{r})w(\mathbf{r}, \omega) = 0, \quad (4)$$

where the elastic constants $D(\mathbf{r}, \omega)$, $\beta(\mathbf{r}, \omega)$, $\gamma(\mathbf{r}, \omega)$, and $a = a(\mathbf{r})$ contain the information of both the hard inclusions ($D_A$, $\beta_A$, $\gamma_A$, $a_A$) and soft matrix ($i\omega D_B(\omega)$, $i\omega \beta_B(\omega)$, $i\omega \gamma_B(\omega)$, $a_B$). Note that once again the term $i\omega$ multiplying the properties of the soft matrix ($B$) arise from the temporal derivative in Eq. (2).

For the SLSM, the temporal part of the elastic constant $\hat{G}_B(t)$, omitting the spatial dependence, can be written as [50, 55]:

$$\hat{G}_B(t) = \left[G_{\infty_B} + (G_{0_B} + G_{\infty_B})e^{-\frac{t}{\tau_{\hat{G}_B}}}\right]\hat{u}(t), \quad (5)$$

where $\hat{u}(t)$ is the unit step function, $\tau_{\hat{G}_B}$ is the relaxation time ($\tau_{\hat{G}_B} = \eta/G_1$ [52], see Fig. 1 (b)), $G_{0_B}$ and $G_{\infty_B}$ are the initial and final states of the elastic constants and are related to $G_{1_B}$, $G_{2_B}$ (Fig. 1 (b)) as $G_{2_B} = G_{\infty_B}$, and $G_{1_B} = G_{0_B} - G_{\infty_B}$. It should be noted, for a linear viscoelastic isotropic case, that $\hat{E}_B(t) = 2(1 + \nu_B)\hat{G}_B(t)$, $\hat{D}_B(t) = \hat{E}_B(t)h^3/12(1 - \nu_B^2)$, $\hat{\beta}_B(t) = \hat{D}_B(t)\nu_B$, and $\hat{\gamma}_B(t) = \hat{D}_B(t)(1 - \nu_B)$.

Applying the temporal Fourier transform to Eq. (5), remembering that $\mathsf{F}\{\hat{u}(t)\} = \pi\delta(\omega) + \frac{1}{i\omega}$, $\omega\delta(\omega) = 0$, and $\mathsf{F}\{e^{-\beta t}u(t)\} = \frac{1}{\beta + i\omega}$, where $\delta(\omega)$ is the Dirac delta function, and $\beta > 0$, results in:

$$G_B(\omega) = \frac{(G_{0_B} - G_{\infty_B})\tau_{\hat{G}_B}}{1 + \omega^2 \tau_{\hat{G}_B}^2} - i\frac{G_{\infty_B} + G_{0_B}\omega^2 \tau_{\hat{G}_B}^2}{\omega(1 + \omega^2 \tau_{\hat{G}_B}^2)}, \quad (6)$$

$\forall \omega \neq 0$, $\lim_{\omega \to 0} G_B(\omega) = (G_{0_B} - G_{\infty_B})\tau_{\hat{G}_B}$, and $E_B(\omega) = 2(1 + \nu_B)G_B(\omega)$, $D_B(\omega) = E_B(\omega)h^3/12(1 - \nu_B^2)$, $\beta_B(\omega) = D_B(\omega)\nu_B$, and $\gamma_B(\omega) = D_B(\omega)(1 - \nu_B)$.

Due to the system periodicity, the Bloch-Floquet theorem [48, 49] implies that:

$$w(\mathbf{r}, \omega) = e^{i\mathbf{k}(\omega) \cdot \mathbf{r}} w_\mathbf{k}(\mathbf{r}), \quad (7)$$

where $w_\mathbf{k}(\mathbf{r})$ is the Bloch wave amplitude and $\mathbf{k} = k_1 \mathbf{e}_1 + k_2 \mathbf{e}_2$.

Expanding $w_\mathbf{k}(\mathbf{r})$ as space Fourier series on the reciprocal space and considering wave propagation in the $xy$ plane ($k_3 = 0$), Eq. (7) can be rewritten as:

$$w(\mathbf{r}, \omega) = e^{i\mathbf{k}(\omega)\cdot\mathbf{r}} \sum_{\mathbf{g}=-\infty}^{+\infty} w(\mathbf{g}) e^{i\mathbf{g}\cdot\mathbf{r}} = \sum_{\mathbf{g}=-\infty}^{+\infty} w(\mathbf{g}) e^{i[\mathbf{k}(\omega)+\mathbf{g}]\cdot\mathbf{r}}, \tag{8}$$

where

$$\mathbf{g} = m\mathbf{b}_1 + n\mathbf{b}_2 = (mb_{1_1} + nb_{2_1})\mathbf{e}_1 + (mb_{1_2} + nb_{2_2})\mathbf{e}_2, \tag{9}$$

with $\mathbf{b}_i = \frac{2\pi}{a}\mathbf{e}_i$ ($i = 1, 2$) are the primitive vectors in reciprocal space for square lattice, $b_{i_{1,2}} = ||\mathbf{b}_{i_{1,2}}||$, and ($m, n \in \mathbb{Z}$). The primitive vectors in the real ($\mathbf{a}_i$) and reciprocal ($\mathbf{b}_i$) spaces for the 2-D viscoelastic PnS plate are illustrated in Fig. 2 ($a, b$), respectively, where $\mathbf{a}_i = a\mathbf{e}_i$.

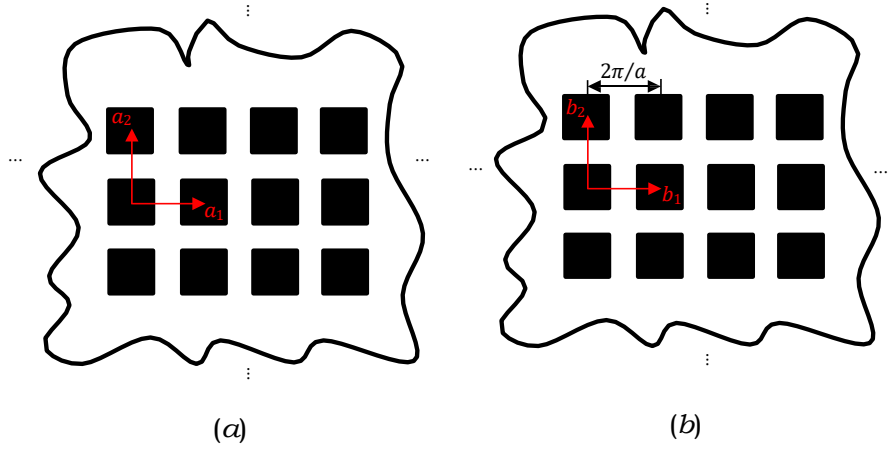

(a)            (b)

Figure 2: Primitive vectors in the (a) real ($\mathbf{a}_i$) and (b) reciprocal ($\mathbf{b}_i$) spaces in the infinite 2-D viscoelastic PnS plate with square hard inclusions in a soft matrix with a square lattice, where $\mathbf{a}_i = a\mathbf{e}_i$, $\mathbf{b}_i = (2\pi/a)\mathbf{e}_i$ ($i = 1, 2$).

The material properties can be expanded as space Fourier series in the reciprocal space as:

$$P(\mathbf{r}, \omega) = \sum_{\bar{\mathbf{g}}=-\infty}^{+\infty} P(\bar{\mathbf{g}}, \omega) e^{i\bar{\mathbf{g}}\cdot\mathbf{r}}, \tag{10}$$

where $P(\mathbf{r}, \omega)$ can be $D(\mathbf{r}, \omega), \beta(\mathbf{r}, \omega)$, or $\gamma(\mathbf{r}, \omega)$, $\bar{\mathbf{g}}$ has the same expression of $\mathbf{g}$, with ($\bar{m}, \bar{n} \in \mathbb{Z}$). Note that $\bar{\mathbf{g}}$ is used instead of $\mathbf{g}$ in order to highlight the difference between the space Fourier series expansions of material properties and the transverse displacement.

The space Fourier series coefficients, $P(\bar{\mathbf{g}}, \omega)$, regarding a square lattice, can be computed by:

$$P(\bar{\mathbf{g}}, \omega) = \frac{1}{S_C} \int P(\mathbf{r}, \omega) e^{-i\bar{\mathbf{g}}\cdot\mathbf{r}} d^2\mathbf{r}, \quad (11)$$

where the integration in Eq. (11) is performed over the unit cell and $S_C = ||\mathbf{a}_1 \times \mathbf{a}_2||$ is the cross-sectional area of the unit cell. Considering a single unit cell of Fig. 1 (a), yields:

$$P(\bar{\mathbf{g}}, \omega) = \bar{P}(\omega)\delta_{\bar{\mathbf{g}}\mathbf{o}} + [P_A - i\omega P_B(\omega)](1 - \delta_{\bar{\mathbf{g}}\mathbf{o}})F(\bar{\mathbf{g}}), \quad (12)$$

where $\delta_{\bar{\mathbf{g}}\mathbf{o}} = 1$, if $\bar{\mathbf{g}} = \mathbf{o}$, or $\delta_{\bar{\mathbf{g}}\mathbf{o}} = 0$, if $\bar{\mathbf{g}} \neq \mathbf{o}$, $\bar{P}(\omega) = \bar{f} P_A + (1 - \bar{f})i\omega P_B(\omega)$, $\bar{f} = \frac{S_A}{S_C}$ is the filling fraction, and $S_A$ is the cross-sectional area of the inclusion. Note that the term $i\omega$ multiplying $P_B(\omega)$ arise from the temporal derivative in Eq. (2). The structure function, $F(\bar{\mathbf{g}})$, depends on the inclusion geometry, and it is defined, for square inclusions, as [57]:

$$F(\bar{\mathbf{g}}) = \bar{f} \operatorname{sinc}(\bar{g}_1 l)\operatorname{sinc}(\bar{g}_2 l), \quad (13)$$

where $\bar{g}_1 = ||\bar{\mathbf{g}}_1|| = \frac{2\pi}{a}m$, $\bar{g}_2 = ||\bar{\mathbf{g}}_2|| = \frac{2\pi}{a}n$, $2l$ is the length of the square inclusions, $\operatorname{sinc}(\bar{g}_i l) = \frac{\sin(\bar{g}_i l)}{\bar{g}_i l}$, $\forall \bar{g}_i \neq 0$, and $\operatorname{sinc}(\bar{g}_i l) = 1$ for $\bar{g}_i l = 0$.

The mass density, $\rho$, can also be expanded as spatial Fourier series in the reciprocal space as:

$$\rho(\mathbf{r}) = \sum_{\bar{\mathbf{g}}=-\infty}^{+\infty} \rho(\bar{\mathbf{g}}) e^{i\bar{\mathbf{g}}\cdot\mathbf{r}}, \quad (14)$$

where $\rho(\mathbf{r})$ is the spatial Fourier series coefficient and it can be computed as:

$$\rho(\bar{\mathbf{g}}) = \frac{1}{S_C} \int \rho(\mathbf{r}) e^{-i\bar{\mathbf{g}}\cdot\mathbf{r}} d^2\mathbf{r}, \quad (15)$$

which considering the unit cell in Fig. 1 (a), yields:

$$\rho(\bar{\mathbf{g}}) = \bar{\rho}\,\delta_{\bar{\mathbf{g}}\mathbf{o}} + (\rho_A - \rho_B)(1 - \delta_{\bar{\mathbf{g}}\mathbf{o}})F(\bar{\mathbf{g}}), \quad (16)$$

where $\bar{\rho} = \bar{f}\rho_A + (1 - \bar{f})\rho_B$.

Substituting Eqs. (8), (10), and (14) in Eq. (4), gives:

$$\sum_{\mathbf{g}=-\infty}^{+\infty}\sum_{\bar{\mathbf{g}}=-\infty}^{+\infty}\{[k_1(\omega)+g_1]^2[k_1(\omega)+g_1+\bar{g}_1]^2 D(\bar{\mathbf{g}},\omega)+[k_2(\omega)+g_2]^2[k_1(\omega)$$
$$+g_1+\bar{g}_1]^2\beta(\bar{\mathbf{g}},\omega)+2[k_1(\omega)+g_1][k_2(\omega)+g_2][k_1(\omega)+g_1+\bar{g}_1][k_2(\omega)+g_2$$
$$+\bar{g}_2]\gamma(\bar{\mathbf{g}},\omega)+[k_2(\omega)+g_2]^2[k_2(\omega)+g_2+\bar{g}_2]^2 D(\bar{\mathbf{g}},\omega)+[k_1(\omega)+g_1]^2[k_2(\omega)$$
$$+g_2+\bar{g}_2]^2\beta(\bar{\mathbf{g}},\omega)-\omega^2 a(\bar{\mathbf{g}})\}w(\mathbf{g})e^{i[\mathbf{k}(\omega)+\mathbf{g}+\bar{\mathbf{g}}]\cdot\mathbf{r}}=0, \quad (17)$$

where $k_i=||\mathbf{k}_i||$ ($i=1,2$). Multiplying Eq. (17) by $e^{-i[\mathbf{k}(\omega)+\tilde{\mathbf{g}}]\cdot\mathbf{r}}/S_C$, with $\tilde{\mathbf{g}}$ has the same expression of $\mathbf{g}$, where ($\tilde{m},\tilde{n}\in\mathbb{Z}$), integrating over the unit cell, yields:

$$\sum_{\mathbf{g}=-\infty}^{+\infty}\sum_{\bar{\mathbf{g}}=-\infty}^{+\infty}\{[k_1(\omega)+g_1]^2[k_1(\omega)+g_1+\bar{g}_1]^2 D(\bar{\mathbf{g}},\omega)+[k_2(\omega)+g_2]^2[k_1(\omega)$$
$$+g_1+\bar{g}_1]^2\beta(\bar{\mathbf{g}},\omega)+2[k_1(\omega)+g_1][k_2(\omega)+g_2][k_1(\omega)+g_1+\bar{g}_1][k_2(\omega)+g_2$$
$$+\bar{g}_2]\gamma(\bar{\mathbf{g}},\omega)+[k_2(\omega)+g_2]^2[k_2(\omega)+g_2+\bar{g}_2]^2 D(\bar{\mathbf{g}},\omega)+[k_1(\omega)+g_1]^2[k_2(\omega)$$
$$+g_2+\bar{g}_2]^2\beta(\bar{\mathbf{g}},\omega)-\omega^2 a(\bar{\mathbf{g}})\}w(\mathbf{g})\frac{1}{S_C}e^{i(\mathbf{g}+\bar{\mathbf{g}}-\tilde{\mathbf{g}})\cdot\mathbf{r}}d^2\mathbf{r}=0. \quad (18)$$

Considering the orthogonal property of the complex exponential series

$$\frac{1}{S_C}\int e^{i(\mathbf{g}+\bar{\mathbf{g}}-\tilde{\mathbf{g}})\cdot\mathbf{r}}d^2\mathbf{r}=\begin{cases}1,&\text{if }\bar{\mathbf{g}}=\tilde{\mathbf{g}}-\mathbf{g}\\0,&\text{otherwise}\end{cases}=\delta_{\bar{\mathbf{g}},\tilde{\mathbf{g}}-\mathbf{g}}, \quad (19)$$

thus Eq. (18) can be rewritten as

$$\sum_{\mathbf{g}=-\infty}^{+\infty}\{[k_1(\omega)+g_1]^2[k_1(\omega)+\tilde{g}_1]^2 D(\tilde{\mathbf{g}}-\mathbf{g},\omega)+[k_2(\omega)+g_2]^2[k_1(\omega)+\tilde{g}_1]^2$$
$$\beta(\tilde{\mathbf{g}}-\mathbf{g},\omega)+2[k_1(\omega)+g_1][k_2(\omega)+g_2][k_1(\omega)+\tilde{g}_1][k_2(\omega)+\tilde{g}_2]\gamma(\tilde{\mathbf{g}}-\mathbf{g},\omega)$$
$$+[k_2(\omega)+g_2]^2[k_2(\omega)+\tilde{g}_2]^2 D(\tilde{\mathbf{g}}-\mathbf{g},\omega)+[k_1(\omega)+g_1]^2[k_2(\omega)+\tilde{g}_2]^2$$
$$\beta(\tilde{\mathbf{g}}-\mathbf{g},\omega)-\omega^2 a(\tilde{\mathbf{g}}-\mathbf{g})\}w(\mathbf{g})=0. \quad (20)$$

The space Fourier series in Eq. (20) should be truncated, in order to obtain a finite system. Choosing $\{m,\tilde{m},n,\tilde{n}\}=[-M,\ldots,M]$, the total number of plane waves is $(2M+1)^2$. Therefore, Eq. (20) can be written in a matrix form

as:
$$[\mathbf{K}(\omega) - \omega^2 \mathbf{M}]\mathbf{w} = \mathbf{0}, \qquad (21)$$

where matrices $\mathbf{K}(\omega)$, $\mathbf{M}$, and the vector $\mathbf{w}$ are described at the following equations. It should be highlighted that matrix $\mathbf{K}(\omega)$ should be computed for each circular frequency value, because of the viscoelasticity of the soft matrix. Thus, the Eq. (21) does not represent a typical generalized eigenvalue problem of $\omega(\mathbf{k})$, and cannot be solved for values of $\mathbf{k}$ scanning the contour of the FIBZ (Fig. 1 (c)).

The matrix $\mathbf{K}(\omega)$ in Eq. (21) is expressed by:

$$\mathbf{K}(\omega) = (\overline{\mathbf{K}} + \mathbf{G})_1^2 \mathbf{D}(\omega)(\overline{\mathbf{K}} + \tilde{\mathbf{G}})_1^2 + (\overline{\mathbf{K}} + \mathbf{G})_2^2 \mathbf{B}(\omega)(\overline{\mathbf{K}} + \tilde{\mathbf{G}})_2^2 + 2(\overline{\mathbf{K}} + \mathbf{G})_1$$
$$(\overline{\mathbf{K}} + \mathbf{G})_2 \mathbf{\Gamma}(\omega)(\overline{\mathbf{K}} + \tilde{\mathbf{G}})_1 (\tilde{\mathbf{K}} + \tilde{\mathbf{G}})_2 + (\overline{\mathbf{K}} + \mathbf{G})_2^2 \mathbf{D}(\omega)(\overline{\mathbf{K}} + \tilde{\mathbf{G}})_2^2 + (\overline{\mathbf{K}} + \mathbf{G})_1^2$$
$$\mathbf{B}(\omega)(\overline{\mathbf{K}} + \tilde{\mathbf{G}})_2^2, \qquad (22)$$

where the matrices $\overline{\mathbf{K}}_i$, $\mathbf{G}_i$, and $\tilde{\mathbf{G}}_i$ ($i = 1, 2$) in Eq. (22) (note that, e.g., $(\overline{\mathbf{K}} + \mathbf{G})_i^2 = \overline{\mathbf{K}}_i^2 + 2\overline{\mathbf{K}}_i \mathbf{G}_i + \mathbf{G}_i^2$, $i = 1, 2$) are given by:

$$\overline{\mathbf{K}}_i = k_i \mathbf{I}, \qquad (23)$$

$$\tilde{\mathbf{G}}_i = \begin{bmatrix} \tilde{g}_i(-M, -M) & 0 & \cdots & 0 \\ 0 & \tilde{g}_i(-M+1, -M+1) & \cdots & 0 \\ \vdots & \vdots & \ddots & \vdots \\ 0 & 0 & \cdots & \tilde{g}_i(M, M) \end{bmatrix}, \qquad (24)$$

and $\mathbf{G}_i = \tilde{\mathbf{G}}_i$.

Matrices $\mathbf{D}(\omega)$, $\mathbf{B}(\omega)$, and $\mathbf{\Gamma}(\omega)$ in Eq. (22) are the matrix form of the space Fourier series coefficients, $D(\tilde{\mathbf{g}} - \mathbf{g}, \omega)$, $B(\tilde{\mathbf{g}} - \mathbf{g}, \omega)$, and $\gamma(\tilde{\mathbf{g}} - \mathbf{g}, \omega)$, respectively. The dependence of $(\tilde{\mathbf{g}} - \mathbf{g})$ is omitted in Eq. (22) and hereafter for brevity. In Eq. (22), the matrices $\mathbf{D}(\omega)$, $\mathbf{B}(\omega)$, and $\mathbf{\Gamma}(\omega)$, in most cases, can be rewritten as $\frac{1}{\mathbf{D}(\tilde{\mathbf{g}}-\mathbf{g},\omega)}^{-1}$, $\frac{1}{\mathbf{B}(\tilde{\mathbf{g}}-\mathbf{g},\omega)}^{-1}$, and $\frac{1}{\mathbf{\Gamma}(\tilde{\mathbf{g}}-\mathbf{g},\omega)}^{-1}$, respectively, i.e., the inverse of the Toeplitz matrices $\frac{1}{\mathbf{D}(\tilde{\mathbf{g}}-\mathbf{g},\omega)}$, $\frac{1}{\mathbf{B}(\tilde{\mathbf{g}}-\mathbf{g},\omega)}$, and $\frac{1}{\mathbf{\Gamma}(\tilde{\mathbf{g}}-\mathbf{g},\omega)}$. The strategy of rewriting the matrices $\mathbf{D}(\omega)$, $\mathbf{B}(\omega)$, and $\mathbf{\Gamma}(\omega)$, usually known as improved plane wave expansion (IPWE) [58] presents higher convergence than the traditional PWE method. However, very recently, Dal Poggetto *et al.* [33] have shown that the IPWE method cannot be applicable for the computation of all matrices, in the context of viscoelastic phononic thick plates when the matrix is hard and inclusions are soft (this is not the case of this study, see Fig. 1 (*a*)). In order to propose a general EPWE formulation, *i.e.*, able to handle also the case proposed by Dal Poggetto *et al.* [33] (hard matrix and soft inclusions), the matrices $\mathbf{D}(\omega)$, $\mathbf{B}(\omega)$, and $\mathbf{\Gamma}(\omega)$ are written in the traditional form hereafter.

The matrix form of the space Fourier series coefficients $\mathbf{D}(\omega)$, $\mathbf{B}(\omega)$, and $\mathbf{\Gamma}(\omega)$ in Eq. (22) can be expressed by:

$$\mathbf{P}(\tilde{\mathbf{g}} - \mathbf{g}, \omega) = \bar{P}\,\mathbf{I} + [P_A + i\omega P_B(\omega)](\mathbf{J} - \mathbf{I}) \circ \mathbf{F}(\tilde{\mathbf{g}} - \mathbf{g}), \tag{25}$$

where $\circ$ represents the Hadamard product, $\mathbf{I}$ is the identity matrix, $\mathbf{J}$ is a matrix of ones, and the matrix form of the structure function Eq. (13), $\mathbf{F}(\tilde{\mathbf{g}} - \mathbf{g})$, is given by:

$$\mathbf{F}(\tilde{\mathbf{g}} - \mathbf{g}) = \begin{bmatrix} F[\tilde{\mathbf{g}}(-M, -M) - \mathbf{g}(-M, -M)] & \cdots & F[\tilde{\mathbf{g}}(M, M) - \mathbf{g}(-M, -M)] \\ \vdots & \ddots & \vdots \\ F[\tilde{\mathbf{g}}(-M, -M) - \mathbf{g}(M, M)] & \cdots & F[\tilde{\mathbf{g}}(M, M) - \mathbf{g}(M, M)] \end{bmatrix}. \tag{26}$$

Matrix $\mathbf{M}$ in Eq. (21) is expressed by:

$$\mathbf{M} = \rho(\tilde{\mathbf{g}} - \mathbf{g}) = \bar{P}\,\mathbf{I} + (P_A + P_B)(\mathbf{J} - \mathbf{I}) \circ \mathbf{F}(\tilde{\mathbf{g}} - \mathbf{g}). \tag{27}$$

Vector $\mathbf{w}$ in Eq. (21) is given by:

$$\mathbf{w}(\mathbf{g}) = \big\{ w[\mathbf{g}(-M, -M)] \quad w[\mathbf{g}(-M+1, -M+1)] \quad \cdots \quad w[\mathbf{g}(M, M)] \big\}^{\mathrm{T}}. \tag{28}$$

From this point on Eq. (21) should be rewritten in the EPWE formulation, *i.e.*, $\mathbf{k}(\omega)$. First, the Bloch wave vector can be computed as $\mathbf{k}(\omega, \bar{\varphi}) = k(\omega)\cos\bar{\varphi}\,\mathbf{e}_1 + k(\omega)\sin\bar{\varphi}\,\mathbf{e}_2$ (see Fig. 1 (*c*)). Thus, the terms $(\bar{\mathbf{K}} + \mathbf{G})^2_{,1}$, $(\bar{\mathbf{K}} + \mathbf{G})^2_{,2}$ and $(\bar{\mathbf{K}} + \mathbf{G})_1(\bar{\mathbf{K}} + \mathbf{G})_2$ of Eq. (22) can be rewritten as:

$$(\bar{\mathbf{K}} + \mathbf{G})_1^2 = \frac{1}{a^2}\{[k(\omega)a\cos(\bar{\varphi})]^2\mathbf{I} + [k(\omega)a\varepsilon_{11}]\mathbf{I} + \varepsilon_{01}\mathbf{I}\}, \quad (29)$$

$$(\bar{\mathbf{K}} + \mathbf{G})_2^2 = \frac{1}{a^2}\{[k(\omega)a\sin(\bar{\varphi})]^2\mathbf{I} + [k(\omega)a\varepsilon_{12}]\mathbf{I} + \varepsilon_{02}\mathbf{I}\}, \quad (30)$$

$$(\bar{\mathbf{K}} + \mathbf{G})_1(\bar{\mathbf{K}} + \mathbf{G})_2 = \frac{1}{a^2}\{[k(\omega)a\cos(\bar{\varphi})\sin(\bar{\varphi})]^2\mathbf{I} + [k(\omega)a\varepsilon_{13}]\mathbf{I} + \varepsilon_{03}\mathbf{I}\}, \quad (31)$$

where

$$\varepsilon_{11} = 2(mb_{1_1} + nb_{2_1})a\cos(\bar{\varphi}), \quad \varepsilon_{01} = [(mb_1 + nb_2)a]^2, \quad (32)$$

$$\varepsilon_{12} = 2(mb_{1_2} + nb_{2_2})a\sin(\bar{\varphi}), \quad \varepsilon_{02} = [(mb_1 + nb_2)a]^2, \quad (33)$$

$$\varepsilon_{13} = [(mb_{1_2} + nb_{2_2})\cos(\bar{\varphi}) + (mb_{1_1} + nb_{2_1})\sin(\bar{\varphi})]a, \quad (34)$$

$$\varepsilon_{03} = (mb_{1_1} + nb_{2_1})(mb_{1_2} + nb_{2_2})a^2. \quad (35)$$

Moreover, the terms $(\bar{\mathbf{K}} + \mathbf{G})_1^2\mathbf{D}(\omega)(\bar{\mathbf{K}} + \tilde{\mathbf{G}})_1^2$, $(\bar{\mathbf{K}} + \mathbf{G})_2^2\mathbf{B}(\omega)(\bar{\mathbf{K}} + \tilde{\mathbf{G}})_1^2$, $2(\bar{\mathbf{K}} + \mathbf{G})_1(\bar{\mathbf{K}} + \mathbf{G})_2\mathbf{\Gamma}(\omega)(\bar{\mathbf{K}} + \tilde{\mathbf{G}})_1(\bar{\mathbf{K}} + \tilde{\mathbf{G}})_2$, $(\bar{\mathbf{K}} + \mathbf{G})_2^2\mathbf{D}(\omega)(\bar{\mathbf{K}} + \tilde{\mathbf{G}})_2^2$, and $(\bar{\mathbf{K}} + \mathbf{G})_1^2\mathbf{B}(\omega)(\bar{\mathbf{K}} + \tilde{\mathbf{G}})_2^2$ of Eq. (22) can be rewritten as:

$$(\bar{\mathbf{K}} + \mathbf{G})_1^2\mathbf{D}(\omega)(\bar{\mathbf{K}} + \tilde{\mathbf{G}})_1^2 = \frac{1}{a^4}\{[k(\omega)a]^4\mathbf{D}_1(\omega,\bar{\varphi}) + [k(\omega)a]^3\mathbf{D}_2(\omega,\bar{\varphi}) + [k(\omega)a]^2\mathbf{D}_3(\omega,\bar{\varphi}) + k(\omega)a\mathbf{D}_4(\omega,\bar{\varphi}) + \mathbf{D}_5(\omega)\},$$

$$\mathbf{D}_1(\omega,\bar{\varphi}) = \cos(\bar{\varphi})^2\mathbf{D}(\omega)\cos(\bar{\varphi})^2, \quad \mathbf{D}_2(\omega,\bar{\varphi}) = \cos(\bar{\varphi})^2\mathbf{D}(\omega)\varepsilon_{11} + \varepsilon_{11}\mathbf{D}(\omega)\cos(\bar{\varphi})^2,$$

$$\mathbf{D}_3(\omega,\bar{\varphi}) = \cos(\bar{\varphi})^2\mathbf{D}(\omega)\varepsilon_{01} + \varepsilon_{11}\mathbf{D}(\omega)\varepsilon_{11} + \varepsilon_{01}\mathbf{D}(\omega)\cos(\bar{\varphi})^2,$$

$$\mathbf{D}_4(\omega,\bar{\varphi}) = \varepsilon_{11}\mathbf{D}(\omega)\varepsilon_{01} + \varepsilon_{01}\mathbf{D}(\omega)\varepsilon_{11}, \quad \mathbf{D}_5(\omega) = \varepsilon_{01}\mathbf{D}(\omega)\varepsilon_{01}, \quad (36)$$

$$(\bar{\mathbf{K}} + \mathbf{G})_2^2\mathbf{B}(\omega)(\bar{\mathbf{K}} + \tilde{\mathbf{G}})_1^2 = \frac{1}{a^4}\{[k(\omega)a]^4\mathbf{B}_1(\omega,\bar{\varphi}) + [k(\omega)a]^3\mathbf{B}_2(\omega,\bar{\varphi}) + [k(\omega)a]^2\mathbf{B}_3(\omega,\bar{\varphi}) + k(\omega)a\mathbf{B}_4(\omega,\bar{\varphi}) + \mathbf{B}_5(\omega)\},$$

$$\mathbf{B}_1(\omega,\bar{\varphi}) = \sin(\bar{\varphi})^2\mathbf{B}(\omega)\cos(\bar{\varphi})^2, \quad \mathbf{B}_2(\omega,\bar{\varphi}) = \sin(\bar{\varphi})^2\mathbf{B}(\omega)\varepsilon_{11} + \varepsilon_{12}\mathbf{B}(\omega)\cos(\bar{\varphi})^2,$$

$$\mathbf{B}_3(\omega,\bar{\varphi}) = \sin(\bar{\varphi})^2\mathbf{B}(\omega)\varepsilon_{01} + \varepsilon_{12}\mathbf{B}(\omega)\varepsilon_{11} + \varepsilon_{01}\mathbf{B}(\omega)\sin(\bar{\varphi})^2,$$

$$\mathbf{B}_4(\omega,\bar{\varphi}) = \varepsilon_{12}\mathbf{B}(\omega)\varepsilon_{01} + \varepsilon_{02}\mathbf{B}(\omega)\varepsilon_{11}, \quad \mathbf{B}_5(\omega) = \varepsilon_{02}\mathbf{B}(\omega)\varepsilon_{01}, \quad (37)$$

$$2(\bar{\mathbf{K}}+\mathbf{G})_1(\bar{\mathbf{K}}+\mathbf{G})_2\mathbf{\Gamma}(\omega)(\bar{\mathbf{K}}+\tilde{\mathbf{G}})_1(\tilde{\mathbf{K}}+\tilde{\mathbf{G}})_2 = \frac{2}{a^4}\{[k(\omega)a]^4\mathbf{\Gamma}_1(\omega,\bar{\varphi})$$
$$+[k(\omega)a]^3\mathbf{\Gamma}_2(\omega,\bar{\varphi})+[k(\omega)a]^2\mathbf{\Gamma}_3(\omega,\bar{\varphi})+k(\omega)a\mathbf{\Gamma}_4(\omega,\bar{\varphi})+\mathbf{\Gamma}_5(\omega)\},$$
$$\mathbf{\Gamma}_1(\omega,\bar{\varphi}) = \cos(\bar{\varphi})\sin(\bar{\varphi})\mathbf{\Gamma}(\omega)\cos(\bar{\varphi})\sin(\bar{\varphi}),$$
$$\mathbf{\Gamma}_2(\omega,\bar{\varphi}) = \cos(\bar{\varphi})\sin(\bar{\varphi})\mathbf{\Gamma}(\omega)\varepsilon_{13} + \varepsilon_{13}\mathbf{\Gamma}(\omega)\cos(\bar{\varphi})\sin(\bar{\varphi}),$$
$$\mathbf{\Gamma}_3(\omega,\bar{\varphi}) = \cos(\bar{\varphi})\sin(\bar{\varphi})\mathbf{\Gamma}(\omega)\varepsilon_{03} + \varepsilon_{13}\mathbf{\Gamma}(\omega)\varepsilon_{13} + \varepsilon_{03}\mathbf{\Gamma}(\omega)\cos(\bar{\varphi})\sin(\bar{\varphi}),$$
$$\mathbf{\Gamma}_4(\omega,\bar{\varphi}) = \varepsilon_{13}\mathbf{\Gamma}(\omega)\varepsilon_{03} + \varepsilon_{03}\mathbf{\Gamma}(\omega)\varepsilon_{13}, \quad \mathbf{\Gamma}_5(\omega) = \varepsilon_{03}\mathbf{\Gamma}(\omega)\varepsilon_{03}, \tag{38}$$

$$(\bar{\mathbf{K}}+\mathbf{G})_2^2\mathbf{D}(\omega)(\bar{\mathbf{K}}+\tilde{\mathbf{G}})_2^2 = \frac{1}{a^4}\{[k(\omega)a]^4\mathbf{D}_6(\omega,\bar{\varphi}) + [k(\omega)a]^3\mathbf{D}_7(\omega,\bar{\varphi})$$
$$+[k(\omega)a]^2\mathbf{D}_8(\omega,\bar{\varphi}) + k(\omega)a\mathbf{D}_9(\omega,\bar{\varphi}) + \mathbf{D}_{10}(\omega)\},$$
$$\mathbf{D}_6(\omega,\bar{\varphi}) = \sin(\bar{\varphi})^2\mathbf{D}(\omega)\sin(\bar{\varphi})^2, \quad \mathbf{D}_7(\omega,\bar{\varphi}) = \sin(\bar{\varphi})^2\mathbf{D}(\omega)\varepsilon_{12} + \varepsilon_{12}\mathbf{D}(\omega)\sin(\bar{\varphi})^2,$$
$$\mathbf{D}_8(\omega,\bar{\varphi}) = \sin(\bar{\varphi})^2\mathbf{D}(\omega)\varepsilon_{02} + \varepsilon_{12}\mathbf{D}(\omega)\varepsilon_{12} + \varepsilon_{02}\mathbf{D}(\omega)\sin(\bar{\varphi})^2,$$
$$\mathbf{D}_9(\omega,\bar{\varphi}) = \varepsilon_{12}\mathbf{D}(\omega)\varepsilon_{02} + \varepsilon_{02}\mathbf{D}(\omega)\varepsilon_{12}, \quad \mathbf{D}_{10}(\omega) = \varepsilon_{02}\mathbf{D}(\omega)\varepsilon_{02}, \tag{39}$$

$$(\bar{\mathbf{K}}+\mathbf{G})_1^2\mathbf{B}(\omega)(\bar{\mathbf{K}}+\tilde{\mathbf{G}})_2^2 = \frac{1}{a^4}\{[k(\omega)a]^4\mathbf{B}_6(\omega,\bar{\varphi}) + [k(\omega)a]^3\mathbf{B}_7(\omega,\bar{\varphi})$$
$$+[k(\omega)a]^2\mathbf{B}_8(\omega,\bar{\varphi}) + k(\omega)a\mathbf{B}_9(\omega,\bar{\varphi}) + \mathbf{B}_{10}(\omega)\},$$
$$\mathbf{B}_6(\omega,\bar{\varphi}) = \cos(\bar{\varphi})^2\mathbf{B}(\omega)\sin(\bar{\varphi})^2, \quad \mathbf{B}_7(\omega,\bar{\varphi}) = \cos(\bar{\varphi})^2\mathbf{B}(\omega)\varepsilon_{12} + \varepsilon_{11}\mathbf{B}(\omega)\sin(\bar{\varphi})^2,$$
$$\mathbf{B}_8(\omega,\bar{\varphi}) = \cos(\bar{\varphi})^2\mathbf{B}(\omega)\varepsilon_{02} + \varepsilon_{11}\mathbf{B}(\omega)\varepsilon_{12} + \varepsilon_{01}\mathbf{B}(\omega)\sin(\bar{\varphi})^2,$$
$$\mathbf{B}_9(\omega,\bar{\varphi}) = \varepsilon_{11}\mathbf{B}(\omega)\varepsilon_{02} + \varepsilon_{01}\mathbf{B}(\omega)\varepsilon_{12}, \quad \mathbf{B}_{10}(\omega) = \varepsilon_{01}\mathbf{B}(\omega)\varepsilon_{02}. \tag{40}$$

Substituting Eqs. (36)-(40) in Eq. (22), Eq. (21) can be rewritten as:
$$\frac{1}{a^4}\{[k(\omega)a]^4\mathbf{A}_1(\omega,\bar{\varphi}) + [k(\omega)a]^3\mathbf{A}_2(\omega,\bar{\varphi}) + [k(\omega)a]^2\mathbf{A}_3(\omega,\bar{\varphi}) + k(\omega)a\mathbf{A}_4(\omega,\bar{\varphi})$$
$$+\mathbf{A}_5(\omega,\bar{\varphi})\}\mathbf{w} = \mathbf{0},$$
$$\mathbf{A}_i(\omega,\bar{\varphi}) = \mathbf{D}_i(\omega,\bar{\varphi}) + \mathbf{B}_i(\omega,\bar{\varphi}) + 2\mathbf{\Gamma}_i(\omega,\bar{\varphi}) + \mathbf{D}_{5+i}(\omega,\bar{\varphi}) + \mathbf{B}_{5+i}(\omega,\bar{\varphi}),$$
$$\mathbf{A}_5(\omega,\bar{\varphi}) = \mathbf{D}_5(\omega,\bar{\varphi}) + \mathbf{B}_5(\omega,\bar{\varphi}) + 2\mathbf{\Gamma}_5(\omega,\bar{\varphi}) + \mathbf{D}_{10}(\omega) + \mathbf{B}_{10}(\omega) - \omega^2 a^4\mathbf{M},$$
$$\tag{41}$$

where $i = 1, \ldots, 4$.

Multiplying Eq. (41) by $\mathbf{A}_1(\omega, \bar{\varphi})^{-1}/a^4$, yields,

$$[\bar{k}^4 \mathbf{I} + \bar{k}^3 \mathbf{A}_6(\omega, \bar{\varphi}) + \bar{k}^2 \mathbf{A}_7(\omega, \bar{\varphi}) + \bar{k} \mathbf{A}_8(\omega, \bar{\varphi}) + \mathbf{A}_9(\omega, \bar{\varphi})]\mathbf{w} = \mathbf{0},$$

$$\mathbf{A}_i(\omega, \bar{\varphi}) = \mathbf{A}_{i-4}(\omega, \bar{\varphi}), \quad \mathbf{A}_9(\omega, \bar{\varphi}) = \mathbf{A}_1^{-1}(\omega, \bar{\varphi})\mathbf{A}_5(\omega, \bar{\varphi}), \qquad (42)$$

where $\bar{k} = k(\omega)a$ and $i = 6, \ldots, 8$.

Thus, it is possible to rewrite the Eq. (42) as a standard eigenvalue problem of $\bar{\mathbf{k}}(\omega, \bar{\varphi})$:

$$\begin{bmatrix} -\mathbf{A}_6(\omega, \bar{\varphi}) & -\mathbf{A}_7(\omega, \bar{\varphi}) & -\mathbf{A}_8(\omega, \bar{\varphi}) & -\mathbf{A}_9(\omega, \bar{\varphi}) \\ \mathbf{I} & \mathbf{0} & \mathbf{0} & \mathbf{0} \\ \mathbf{0} & \mathbf{I} & \mathbf{0} & \mathbf{0} \\ \mathbf{0} & \mathbf{0} & \mathbf{I} & \mathbf{0} \end{bmatrix} \begin{bmatrix} \bar{k}^3 \mathbf{w} \\ \bar{k}^2 \mathbf{w} \\ \bar{k} \mathbf{w} \\ \mathbf{w} \end{bmatrix} = \bar{k} \begin{bmatrix} \bar{k}^3 \mathbf{w} \\ \bar{k}^2 \mathbf{w} \\ \bar{k} \mathbf{w} \\ \mathbf{w} \end{bmatrix}.$$

(43)

For a given frequency $\omega$ and an azimuth angle $\bar{\varphi}$ of $\mathbf{k}(\omega, \bar{\varphi})$, there are $4(2M+1)^2$ eigenvalues $\bar{k}$. It should be highlighted that the complex values of $\mathbf{k}$, *i.e.*, the evanescent wave behavior, are obtained only by EPWE, which is not yielded when using the PWE and IPWE.

## 3. Simulated Example

The viscoelastic phononic thin plate geometry and material properties used for the simulation are presented in Table 1. It is composed by steel square inclusions (A) in an epoxy matrix (B) with a square lattice (see Fig. 1 (*a*)).

It should be highlighted that the thin plate theory (*i.e.*, $||\mathbf{k}||h \ll 1$, $h/a \ll 1$, [17] or $h < \lambda_{\min}/6$ [60], where $\lambda_{\min} = \min\{\sqrt[4]{||D_B(\omega_{\max})||/\rho_B h}\, 2\pi/f_{\max}, \sqrt[4]{D_A/\rho_A h}\, 2\pi/f_{\max}\}$) is fulfilled, since $h/a = 0.005$, $\lambda_{\min} = 0.0487$ m, and $h < 0.0081$ ($\lambda_{\min}/6$) m.

The complex band structure calculated by the improved EPWE formulation (*i.e.*, considering the inverse of the Toeplitz matrices $\frac{1}{P(\mathbf{g}-\mathbf{g},\omega)}$, see Eq. (25)) is ordered using the model assurance criterion (MAC) [61]. The MAC estimates the correlation among the Bloch wave mode shapes obtained by the EPWE approach. Furthermore, the integers $m$, $\tilde{m}$, $n$, and $\tilde{n}$ are limited to the interval $[-3, 3]$, *i.e.*, 49 plane waves were used for the spatial Fourier series expansion, in order to reduce the computational time. Moreover, we underline that the convergence of spatial Fourier series is not investigated in this study, since it depends on frequency (because of the viscoelastic effect) and filling fraction.

Table 1: Geometry and material properties of the viscoelastic phononic thin plate composed by steel square inclusions (A) [59] in an epoxy matrix (B) [50, 51] with a square lattice.

| Geometry/Property | Value |
| --- | --- |
| Lattice parameter ($a$) | 0.11 m |
| Plate thickness ($h$) | $0.005a$ m |
| Filling fraction ($\bar{f}$) | 0.1 |
| Mass density ($\rho_A$, $\rho_B$) | $7.835 \times 10^3$ kg/m$^3$, $1.18 \times 10^3$ kg/m$^3$ |
| Young's modulus ($E_A$, $E_{0_B}$) | $210.3 \times 10^9$ Pa, $3.4918 \times 10^9$ Pa |
| Shear modulus ($G_A$, $G_{0_B}$) | $81.65 \times 10^9$ Pa, $1.58 \times 10^9$ Pa |
| Poisson's ratio ($\nu_A$, $\nu_B$) | 0.2878 , 0.105 |

It should also be mentioned that the complex band structures in this study are computed along the ΓX direction (*i.e.*, $\bar{\varphi} = 0$) since this is a commonly assumed direction to analyze the evanescent behavior of wave dispersion in PnSs using a **k**($\omega$) approach [30, 33, 62]. Unlike in the **k**($\omega$) approach, the wave dispersion relations can be obtained scanning the contour of the FIBZ using an $\omega$(**k**) approach [33]. In addition, the influence of different viscosities and relaxation times on the complex band structure using the EPWE and the SLSM is not investigated, since it was recently reported by Oliveira *et al.* [38] for 1-D PnSs.

Figure 3 shows the complex band structure of the phononic thin plate without viscoelastic effects. Hereafter, a discretization of 0.1 Hz is considered when using the EPWE approach. The real part of the normalized wave number (**k**$a/2\pi$) is illustrated in the Fig. 3 (*a*), the normalized frequency is given by $\Omega = \omega a/2\pi c_{t_B}$, where $c_{t_B} = \sqrt{G_{0_B}/\rho_B}$ is the transverse shear wave velocity in the epoxy matrix, and the band structure is computed by the improved PWE (black circles) [50] and the proposed EPWE (coloured points) approaches.

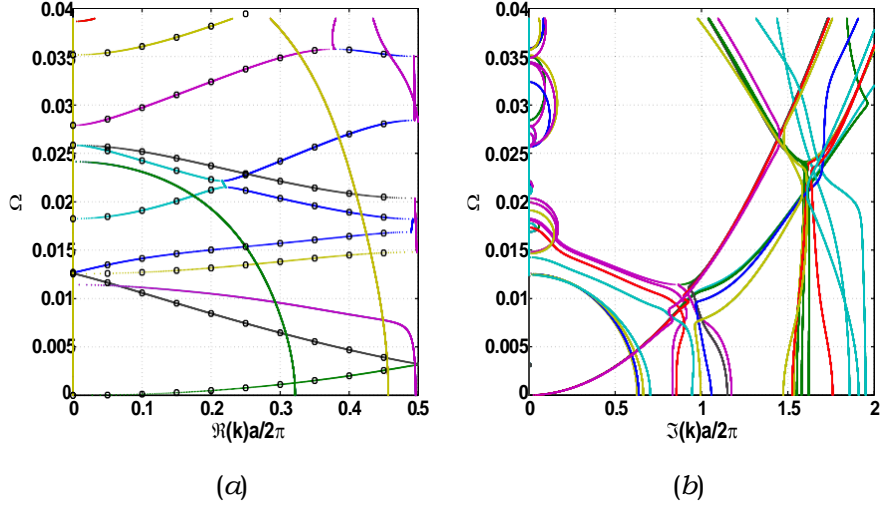

(a)            (b)

Figure 3: Complex band structure of the phononic thin plate with steel inclusions in a epoxy matrix (without viscoelasticity) computed by (a) PWE (black circles) and (a) – (b) EPWE (coloured points) approaches, where $\Omega = \omega a/2\pi c_{t_B}$ and $c_{t_B} = \sqrt{G_{0_B}/\rho_B}$.

A good matching between the PWE and EPWE methods is observed in Fig. 3 (a), however, some modes captured by the proposed EPWE are not obtained directly by the PWE, because these modes are complex and the PWE only identifies pure propagating (real) modes [31]. The evanescent Bloch waves cannot propagate within the phononic thin plate, since their amplitudes decrease with an attenuation distance determined by the imaginary part of **k** in Fig. 3 (b). In Fig. 3 (b), only the positive evanescent wave modes (symmetric negative wave modes exist) are illustrated until $\Im\{\mathbf{k}\}a/2\pi = 2$ (higher values exist). This typical evanescent wave dispersion behavior in Fig. 3 (b) has already been reported for PnSs without viscoelastic effect [30–32].

Figure 4 illustrates the complex band structure of the phononic thin plate with viscoelasticity, considering the SLSM (Fig. 1 (b)), $\tau_{G_B} = 1 \times 10^{-4}$ s, and $G_{\infty_B} = G_{0_B}/5$.

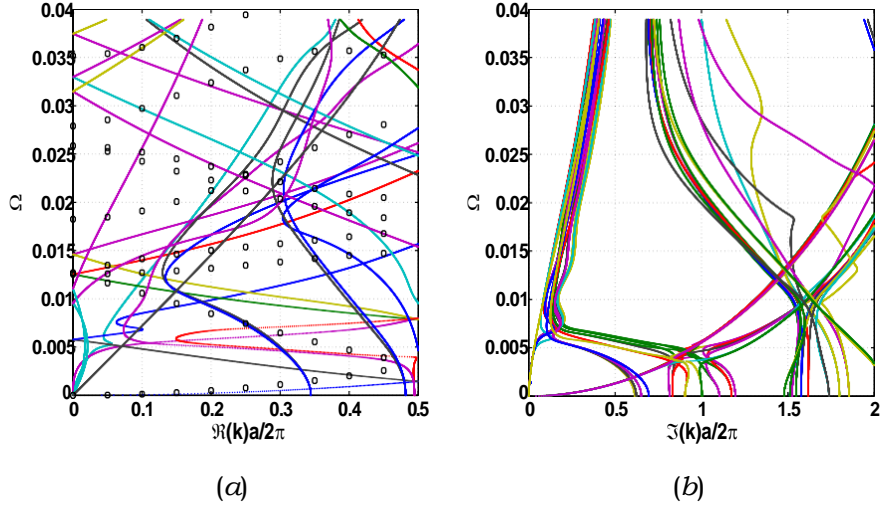

(a)            (b)

Figure 4: Complex band structure of the phononic thin plate with steel inclusions in a epoxy matrix (with viscoelasticity considering the SLSM, $\tau_{G_B} = 1 \times 10^{-4}$ s, and $G_{\infty\ B} = G_{0_B}/5$) computed by (a) PWE (black circles, without viscoelasticity) and (a) – (b) EPWE (coloured points) approaches, where $\Omega = \omega a/2\pi c_{t_B}$ and $c_{t_B} = \overline{G_{0_B}/\rho_B}$.

It can be observed that the PWE (black circles) cannot obtain the correct real part of wave modes (Fig. 4 (a)), since all Bloch wave modes are complexes because of the viscoelasticity, even for lower frequencies. Moreover, the unit cell wave attenuation (*i.e.*, $\Im\{\mathbf{k}\}a$) is reported in the Fig. 4 (b). In Fig. 4 (b), it should be mentioned that some wave modes are shifted from the origin, which is a typical behavior of the viscoelastic PnSs [40].

In Fig. 5, it is shown the imaginary part zoom of the complex band structure of the phononic thin plate without (a) and with (b) viscoelasticity (considering the SLSM, $\tau_{G_B} = 1 \times 10^{-4}$ s, and $G_{\infty\ B} = G_{0_B}/5$) computed by the EPWE (coloured points) approach. It can be seen in (b) that the wave modes are shifted from the origin.

Another interesting issue related to the complex band structure behavior is the influence of the filling fraction ($\bar{f}$). The effect of $\bar{f}$ on the Bragg scattering band gap formation (considering only the propagating waves) is well-known [17, 46, 63]. However, its influence on the evanescent behavior of viscoelastic phononic thin plates has not been reported yet. First, it is illustrated in Fig. 6 the complex band structure (only the second mode is shown, since it is the least attenuated wave mode in Fig. 4 (b)) of the phononic thin plate with viscoelasticity, considering the SLSM (Fig. 1 (b)), $\tau_{\hat{G}_B} = 1 \times 10^{-4}$ s, and $G_{\infty_B} = G_{0_B}/5$.

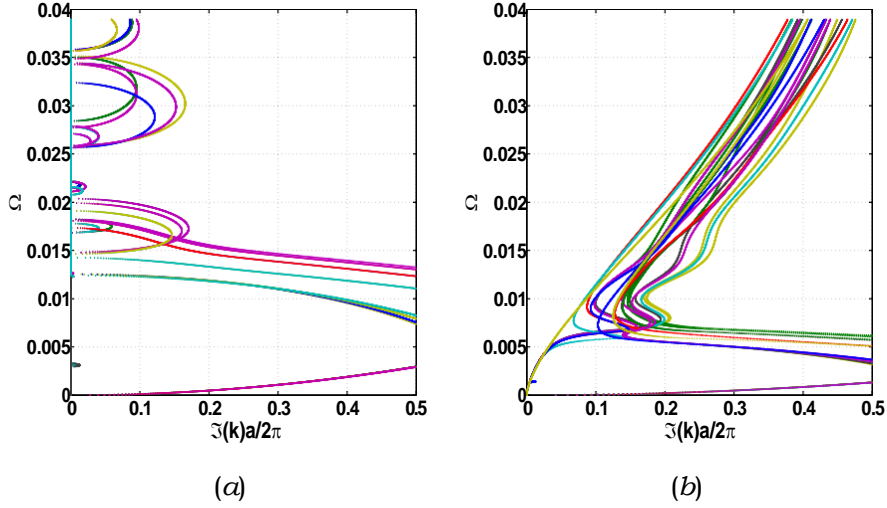

Figure 5: Imaginary part zoom of the complex band structure of the phononic thin plate with steel inclusions in a epoxy matrix, (a) without and (b) with viscoelasticity (considering the SLSM, $\tau_{\hat{G}_B} = 1 \times 10^{-4}$ s, and $G_{\infty_B} = G_{0_B}/5$) computed by EPWE (coloured points) approach, where $\Omega = \omega a/2\pi c_{t_B}$ and $c_{t_B} = \overline{\sqrt{G_{0_B}/\rho_B}}$.

It should be highlighted that wave modes corresponding to the branch with the smallest imaginary part of **k** (least attenuated waves, *i.e.*, the second wave mode in Fig. 4 (b)) contribute the most to the evanescent behavior [33, 64]. In Fig. 6 (b), it can be observed a frequency region with a peak of unit cell wave attenuation ($\Im\{\mathbf{k}\}a/2\pi = 0.1696$) around $\Omega = 0.0071$, considering a fixed value of $\bar{f} = 0.1$. In Fig. 6 (b), it can also be seen that this wave mode is shifted from the origin and it increases with frequency.

In Fig. 7, it is shown the imaginary part ($||\Im\{\mathbf{k}\}||a/2\pi$, color bar) of the complex band structure (only the second mode is shown) of the phononic thin plate with viscoelasticity considering the SLSM, $\tau_{\hat{G}_B} = 1 \times 10^{-4}$ s, $G_{\infty_B} =$

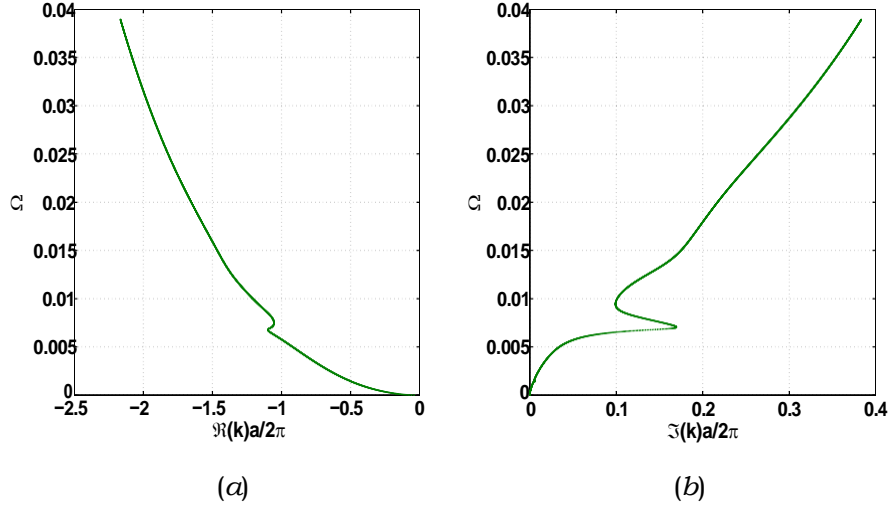

(a)              (b)

Figure 6: Real (*a*) and imaginary (*b*) parts of the complex band structure (only the second mode is shown) of the phononic thin plate with steel inclusions in an epoxy matrix (with viscoelasticity considering the SLSM, $\tau_{G_B} = 1 \times 10^{-4}$ s, and $G_{\infty_B} = G_{0_B}/5$) computed by EPWE, where $\Omega = \omega a/2\pi c_{t_B}$ and $c_{t_B} = \sqrt{G_{0_B}/\rho_B}$.

$G_{0_B}/5$, ranging with filling fraction ($0.01 \leq \bar{f} \leq 0.99$), and $\Omega$. It can be seen that the wave attenuation increases with $\Omega$ for lower values of filling fraction, which is a typical behavior of viscoelastic periodic systems [40]. Moreover, there are interesting regions of filling fraction in lower ($0.78 \leq \bar{f} \leq 0.94$, $\Omega < 0.006$) and higher ($0.10 \leq \bar{f} \leq 0.74$, $\Omega \geq 0.006$) frequencies that present high wave attenuation. The highest wave attenuation region is found around $\bar{f} = 0.37$ for higher frequencies ($0.03 \leq \Omega \leq 0.039$).

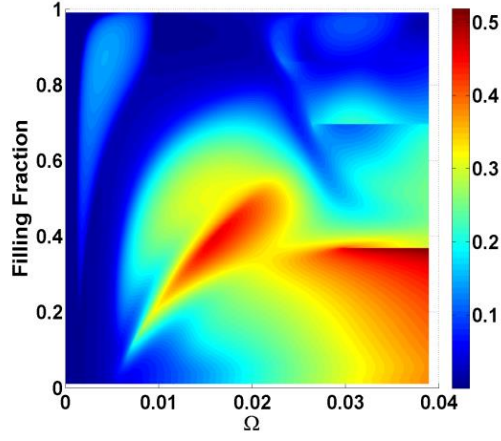

Figure 7: Imaginary part ($||S\{\mathbf{k}\}||a/2\pi$, color bar) of the complex band structure (considering only the second mode) of the phononic thin plate with steel inclusions in a epoxy matrix (with viscoelasticity considering the SLSM, $\tau_{\hat{G}_B} = 1 \times 10^{-4}$ s, and $G_{\infty_B} = G_{0_B}/5$) computed by EPWE ranging with filling fraction ($0.01 \leq \bar{f} \leq 0.99$) and $\Omega = \omega a / 2\pi c_{t_B}$, where $c_{t_B} = \overline{G_{0_B}/\rho_B}$.

## 4. Conclusions

The EPWE formulation is proposed to compute the complex band structure of a viscoelastic phononic thin plate, considering the SLSM. This $\mathbf{k}(\omega)$ approach is important because the evanescent behavior of wave dispersion in viscoelastic periodic structures still has knowledge gaps and it has a considerably lower computational cost compared to other approaches, such as the finite element method. The traditional PWE cannot compute the correct Bloch wave modes of the viscoelastic phononic thin plate, even for lower frequencies. Only the proposed EPWE can compute the correct complex Bloch wave modes and also the unit cell wave attenuation for viscoelastic periodic thin plate structures considering the SLSM.

A good agreement between the PWE and EPWE methods is observed for the case without viscoelasticity, however, some wave modes captured by the proposed EPWE are not obtained directly by the PWE, because these wave modes are complex and the PWE only identifies pure propagating wave modes. The viscoelastic effet influences the unit cell wave attenuation in periodic thin plate structures. The higher unit cell wave attenuation of the viscoelastic phononic thin plate is found around $\bar{f}$ = 0.37 (0.03 ≤ Ω ≤ 0.039), considering the least attenuated wave mode. Furthermore, the **k**($\omega$) approach proposed in this study can be extended to other plate theories and also to different viscoelastic models.

**Acknowledgments**

EJPM and JMCDS thank the Federal Institute of Maranhão (IFMA), Brazilian funding agencies CAPES (Finance Code 001), CNPq (grants 313620/2018, 403234/2021-2, 405638/2022-1, and 300166/2022-2), FAPEMA (grants 07168/22 and 09683/22), and FAPESP (grant 2018/15894-0). VFDP and NMP are supported by the EU H2020 FET Open "Boheme" grant No. 863179.

**References**

[1] M. M. Sigalas, E. N. Economou, Elastic and acoustic wave band structure, Journal of Sound and Vibration 158 (2) (1992) 377–382. doi:10.1016/0022-460X(92)90059-7.

[2] M. Maldovan, Sound and heat revolutions in phononics, Nature 503 (2013) 209–217. doi:10.1038/nature12608.

[3] W. Zhou, Z. Chen, Y. Chen, W. Chen, C. W. Lima, J. N. Reddy, Mathematical modelling of phononic nanoplate and its size-dependent dispersion and topological properties, Applied Mathematical Modelling 88 (2020) 774–790. doi:10.1016/j.apm.2020.07.008.

[4] N. Gao, J. H. Wu, L. Yu, Low frequency band gaps below 10 Hz in radial flexible elastic metamaterial plate, Journal of Physics D: Applied Physics 49 (435501) (2016) 1–9. doi:10.1088/0022-3727/49/43/435501.

[5] A. T. Fabro, D. Beli, N. S. Ferguson, J. R. F. Arruda, B. R. Mace, Wave and vibration analysis of elastic metamaterial and phononic crystal beams with slowly varying properties, Wave Motion 103 (2021) 102728. doi:10.1016/j.wavemoti.2021.102728.

[6] R. G. Salsa Jr., T. P. Sales, D. A. Rade, Optimization of vibration band gaps in damped lattice metamaterials, Latin American Journal of Solids and Structures 20 (6) (2022) e493. doi:10.1590/1679-78257486.

[7] M. Morvaridi, G. Carta, M. Brun, Platonic crystal with low-frequency locally-resonant spiral structures: wave trapping, transmission amplification, shielding and edge waves, Journal of the Mechanics and Physics of Solids 121 (2018) 496–516. doi:10.1016/j.jmps.2018.08.017.

[8] S.-H. Jo, B. D. Youn, Enhanced ultrasonic wave generation using energy-localized behaviors of phononic crystals, International Journal of Mechanical Sciences 228 (2022) 107483. doi:10.1016/j.ijmecsci.2022.107483.

[9] D. P. Vasconcellos, R. S. Cruz, J. C. M. Fernandes, M. Silveira, Vibration attenuation and energy harvesting in metastructures with nonlinear absorbers conserving mass and strain energy, The European Physical Journal Special Topics 231 (2022) 1393–1401. doi:10.1140/epjs/s11734-022-00489-7.

[10] C. C. Claeys, K. Vergote, P. Sas, W. Desmet, On the potential of tuned resonators to obtain low-frequency vibrational stop bands in periodic panels, Journal of Sound and Vibration 332 (2013) 1418–1436. doi:10.1016/j.jsv.2012.09.047.

[11] K. M. Ho, C. K. Cheng, Z. Yang, X. X. Zhang, P. Sheng, Broadband locally resonant sonic shields, Applied Physics Letters 83 (26) (2003) 5566–5568. doi:10.1063/1.1637152.

[12] J. Yang, J. S. Lee, Y. Y. Kim, Metaporous layer to overcome the thickness constraint for broadband sound absorption, Journal of Applied Physics 117 (17) (2015) 174903. doi:10.1063/1.4919844.

[13] L. Y. M. Sampaio, P. C. M. Cerântola, L. P. R. Oliveira, Lightweight decorated membranes as an aesthetic solution for sound insulation panels, Journal of Sound and Vibration 532 (2022) 116971. doi:10.1016/j.jsv.2022.116971.

[14] E. J. P. Miranda Jr., J. M. C. Dos Santos, Flexural wave band gaps in multi-resonator elastic metamaterial timoshenko beams, Wave Motion 91 (2019) 102391. doi:10.1016/j.wavemoti.2019.102391.

[15] M. M. Sigalas, E. N. Economou, Elastic waves in plates with periodically placed inclusions, Journal of Applied Physics 75 (1994) 2845–2850. doi:10.1063/1.356177.

[16] M. S. Kushwaha, Classical band structure of periodic elastic composites, International Journal of Modern Physics B 10 (9) (1996) 977–1094. doi:10.1142/S0217979296000398.

[17] Z. J. Yao, G. L. Yu, Y. S. Wang, Z. F. Shi, Propagation of bending waves in phononic crystal thin plates with a point defect, International Journal of Solids and Structures 46 (13) (2009) 2571–2576. doi:10.1016/j.ijsolstr.2009.02.002.

[18] Z. B. Cheng, Y. G. Xu, L. L. Zhang, Analysis of flexural wave bandgaps in periodic plate structures using differential quadrature element method, International Journal of Mechanical Sciences 100 (2015) 112–125. doi:10.1016/j.ijmecsci.2015.06.014.

[19] X. K. Han, Z. Zhang, Topological optimization of phononic crystal thin plate by a genetic algorithm, Scientific Reports 9 (2019) 8331. doi:10.1038/s41598-019-44850-8.

[20] V. F. Dal Poggetto, J. R. F. Arruda, Widening wave band gaps of periodic plates via shape optimization using spatial Fourier coefficients, Mechanical Systems and Signal Processing 147 (2021) 107098. doi:10.1016/j.ymssp.2020.107098.

[21] G. Yan, S. Yao, Y. Li, Propagation of elastic waves in metamaterial plates with various lattices for low-frequency vibration attenuation, Journal of Sound and Vibration 536 (2022) 117140. doi:10.1016/j.jsv.2022.117140.

[22] V. F. Dal Poggetto, A. L. Serpa, Elastic wave band gaps in a three-dimensional periodic metamaterial using the plane wave expansion method, International Journal of Mechanical Sciences 184 (2020) 105841. doi:10.1016/j.ijmecsci.2020.105841.

[23] L. H. M. S. Ribeiro, V. F. Dal Poggetto, J. R. F. Arruda, Robust optimization of attenuation bands of three-dimensional periodic frame structures, Acta Mechanica 233 (2022) 455–475. doi:10.1007/s00707-021-03118-x.

[24] E. J. P. Miranda Jr., E. D. Nobrega, S. F. Rodrigues, C. Aranas Jr., J. M. C. Dos Santos, Wave attenuation in elastic metamaterial thick

plates: Analytical, numerical and experimental investigations, International Journal of Solids and Structures 204-205 (2020) 138–152. doi:10.1016/j.ijsolstr.2020.08.002.

[25] A. B. Movchan, N. V. Movchan, R. C. McPhedran, Bloch-Floquet bending waves in perforated thin plates, Proceedings of the Royal Society A: Mathematical, Physical and Engineering Sciences 463 (2086) (2007) 2505–2518. doi:10.1098/rspa.2007.1886.

[26] N. V. Movchan, R. C. McPhedran, A. B. Movchan, C. G. Poulton, Wave scattering by platonic grating stacks, Proceedings of the Royal Society A: Mathematical, Physical and Engineering Sciences 465 (2111) (2011) 3383–3400. doi:10.1098/rspa.2009.0301.

[27] C. G. Poulton, R. C. McPhedran, N. V. Movchan, A. B. Movchan, Convergence properties and flat bands in platonic crystal band structures using the multipole formulation, Waves in Random and Complex Media 20 (4) (2010) 702–716. doi:10.1080/17455030903203140.

[28] N. V. Movchan, R. C. McPhedran, A. B. Movchan, Flexural waves in structured elastic plates: Mindlin versus bi-harmonic models, Proceedings of the Royal Society A: Mathematical, Physical and Engineering Sciences 467 (2127) (2011) 869–880. doi:10.1098/rspa.2010.0375.

[29] L. Brillouin, Wave propagation in periodic structures, Dover Publications, New York, 1946.

[30] V. Laude, Y. Achaoui, S. Benchabane, A. Khelif, Evanescent Bloch waves and the complex band structure of phononic crystals, Physical Review B 80 (092301) (2009) 1–4. doi:10.1103/PhysRevB.80.092301.

[31] E. J. P. Miranda Jr., S. F. Rodrigues, J. M. C. Dos Santos, Complex dispersion diagram and evanescent modes in piezomagnetic phononic structures, Solid State Communications 346 (2022) 114697. doi:10.1016/j.ssc.2022.114697.

[32] E. J. P. Miranda Jr., J. M. C. Dos Santos, Wave attenuation in 1-3 phononic structures with lead-free piezoelectric ceramic inclusions, Physica B: Physics of Condensed Matter 631 (2022) 413642. doi:10.1016/j.physb.2021.413642.

[33] V. F. Dal Poggetto, E. J. P. Miranda Jr., J. M. C. Dos Santos, N. M. Pugno, Wave attenuation in viscoelastic hierarchical plates, International Journal of Mechanical Sciences 236 (2022) 107763. doi:10.1016/j.ijmecsci.2022.107763.

[34] W. J. Parnell, R. De Pascalis, Soft metamaterials with dynamic viscoelastic functionality tuned by pre-deformation, Philosophical Transactions of the Royal Society A 377 (2144) (2019) 1–26. doi:10.1098/rsta.2018.0072.

[35] M. I. Mustafa, Viscoelastic Timoshenko beams with variable-exponent nonlinearity, Journal of Mathematical Analysis and Applications 516 (2022) 126520. doi:10.1016/j.jmaa.2022.126520.

[36] B. M. R. Calsavara, E. H. G. Tavares, M. A. J. Silva, Exponential stability for a thermo-viscoelastic Timoshenko system with fading memory, Journal of Mathematical Analysis and Applications 512 (2022) 126147. doi:10.1016/j.jmaa.2022.126147.

[37] L. F. C. Schalcher, J. M. C. Dos Santos, E. J. P. Miranda Jr., Extended plane wave expansion formulation for 1-D viscoelastic phononic crystals, Partial Differential Equations in Applied Mathematics 7 (2023) 100489. doi:10.1016/j.padiff.2023.100489.

[38] V. B. S. Oliveira, L. F. C. Schalcher, J. M. C. Dos Santos, E. J. P. Miranda Jr., Wave attenuation in 1-D viscoelastic phononic crystal rods using different polymers, Materials Research 26 (Suppl.1) (2023) e20220534. doi:10.1590/1980-5373-MR-2022-0534.

[39] M. I. Hussein, Theory of damped Bloch waves in elastic media, Physical Review B 80 (21) (212301). doi:10.1103/PhysRevB.80.212301.

[40] R. P. Moiseyenko, V. Laude, Material loss influence on the complex band structure and group velocity in phononic crystals, Physical Review B 83 (2011) 064301. doi:10.1103/PhysRevB.83.064301.

[41] G. Kirchhoff, Über das gleichgewicht und die bewegung einer elastischen scheibe, Journal für die reine and angewandtle Mathematik 40 (1850) 51–88.

[42] A. E. H. Love, The small free vibrations and deformation of a thin elastic shell, Philosophical Transactions of the Royal Society 179 (1888) 491–546.

[43] Y. Xiao, J. Wen, X. Wen, Flexural wave band gaps in locally resonant thin plates with periodically attached spring-mass resonators, Journal of Physics D: Applied Physics 45 (19) (2012) 1–12. doi:10.1088/0022-3727/45/19/195401.

[44] E. J. P. Miranda Jr., E. D. Nobrega, A. H. R. Ferreira, J. M. C. Dos Santos, Flexural wave band gaps in a multi-resonator elastic metamaterial plate using Kirchhoff-Love theory, Mechanical Systems and Signal Processing 116 (2019) 480–504. doi:10.1016/j.ymssp.2018.06.059.

[45] E. J. P. Miranda Jr., S. F. Rodrigues, C. Aranas Jr., J. M. C. Dos Santos, Plane wave expansion and extended plane wave expansion formulations for Mindlin-Reissner elastic metamaterial thick plates, Journal of Mathematical Analysis and Applications 505 (2022) 125503. doi:10.1016/j.jmaa.2021.125503.

[46] E. J. P. Miranda Jr., J. M. C. Dos Santos, Evanescent Bloch waves and complex band structure in magnetoelectroelastic phononic crystals, Mechanical Systems and Signal Processing 112 (2018) 280–304. doi:10.1016/j.ymssp.2018.04.034.

[47] Y. C. Hsue, A. J. Freeman, Extended plane-wave expansion method in three-dimensional anisotropic photonic crystals, Physical Review B 72 (195118) (2005) 1–10. doi:10.1103/PhysRevB.72.195118.

[48] G. Floquet, Sur les équations différentielles linéaires à coefficients périodiques, Annales scientifiques de l'École Normale Supérieure 12 (1883) 47–88.

[49] F. Bloch, Über die quantenmechanik der electron in kristallgittern, Zeitschrift für Physik 52 (1928) 555–600.

[50] Y. P. Zhao, P. J. Wei, The band gap of 1D viscoelastic phononic crystal, Computational Materials Science 46 (2009) 603–606. doi:10.1016/j.commatsci.2009.03.040.

[51] P. J. Wei, Y. P. Zhao, The influence of viscosity on band gaps of 2D phononic crystal, Mechanics of Advanced Materials and Structures 17 (2010) 383–392. doi:10.1080/15376494.2010.483320.

[52] R. Lakes, Viscoelastic Materials, Cambridge University Press, New York, 2009.

[53] C.-Y. Li, Alternative form of standard linear solid model for characterizing stress relaxation and creep: including a novel parameter for quantifying the ratio of fluids to solids of a viscoelastic solid, Frontiers in Materials 7 (11) (2020) 1–11. doi:10.3389/fmats.2020.00011.

[54] M. E. Gurtin, E. Sternberg, On the linear theory of viscoelasticity, Archive for Rational Mechanics and Analysis 11 (1962) 291–356. doi:10.1007/BF00253942.

[55] F.-L. Li, C. Zhang, Y.-S. Wang, Analysis of the effects of viscosity on the sh-wave band-gaps of 2D viscoelastic phononic crystals by Dirichlet-to-Neumann map method, International Journal of Mechanical Sciences 195 (2021) 106225. doi:10.1016/j.ijmecsci.2020.106225.

[56] N. W. Tschoegl, W. G. Knauss, I. Emri, Poisson's ratio in linear viscoelasticity - A critical review, Mechanics of Time-Dependent Materials 6 (2002) 3–51. doi:10.1023/A:1014411503170.

[57] J. O. Vasseur, B. Djafari-Rouhani, L. Dobrzynski, M. S. Kushwaha, P. Halevi, Complete acoustic band gaps in periodic fibre reinforced composite materials: the carbodepoxy composite and some metallic systems, Journal of Physics: Condensed Matter 6 (1994) 8759–8770. doi:10.1088/0953-8984/6/42/008.

[58] Y. Cao, Z. Hou, Y. Liu, Convergence problem of plane-wave expansion method for phononic crystals, Physics Letters A 327 (2004) 247–253. doi:10.1016/j.physleta.2004.05.030.

[59] S. A. Kim, W. L. Johnson, Elastic constants and internal friction of martensitic steel, ferritic-pearlitic steel, and $\alpha$-iron, Materials Science and Engineering A 452-453 (2007) 633–639. doi:10.1016/j.msea.2006.11.147.

[60] F. Fahy, P. Gardonio, Sound and structural vibration: Radiation, transmission and response, Academic Press, 2007.

[61] J.-M. Mencik, On the low- and mid-frequency forced response of elastic structures using wave finite elements with one-dimensional propagation, Computers and Structures 88 (2010) 674–689. doi:10.1016/j.compstruc.2010.02.006.

[62] Y. Chen, X. Huang, G. Sun, X. Yan, G. Li, Maximizing spatial decay of evanescent waves in phononic crystals by topology optimization, Computers and Structures 182 (2017) 430–447. doi:10.1016/j.compstruc.2017.01.001.

[63] M. S. Kushwaha, P. Halevi, G. Martínez, L. Dobrzynski, B. Djafari-Rouhani, Theory of acoustic band structure of periodic elastic composites, Physical Review B 49 (4) (1994) 2313–2322. doi:10.1103/PhysRevB.49.2313.

[64] Y. Xiao, J. Wen, X. Wen, Broadband locally resonant beams containing multiple periodic arrays of attached resonators, Physics Letters A 376 (16) (2012) 1384–1390. doi:10.1016/j.physleta.2012.02.059.